\documentclass[aps,prl,twocolumn,superscriptaddress,showkeys,floatfix,pifont,citesort,epsf]{revtex4}
\usepackage{graphicx}
\usepackage{amssymb}
\usepackage{amsmath}
\usepackage{dcolumn}
\usepackage{color}
\usepackage{multirow}

\begin{document}

\title{Spin-Peierls transition in TiPO$_4$}

\author{J. M. Law}
\email{j.law@fkf.mpg.de}
\affiliation{Max Planck Institut f\"ur Festk\"orperforschung,
Heisenbergstr. 1, D-70569 Stuttgart, Germany}
\affiliation{Physics Department, Loughborough University, Leicestershire, UK}

\author{C. Hoch}
\affiliation{Max Planck Institut f\"ur Festk\"orperforschung,
Heisenbergstr. 1, D-70569 Stuttgart, Germany}

\author{R. Glaum}
\affiliation{Institut f$\ddot{u}$r Anorganische Chemie, Universit$\ddot{a}$t Bonn,  53121 Bonn, Germany}

\author{I. Heinmaa}
\affiliation{National Institute of Chemical Physics And Biophysics,  12618 Tallinn, Estonia}

\author{R. Stern}
\affiliation{National Institute of Chemical Physics And Biophysics,  12618 Tallinn, Estonia}

\author{J. Kang}
\affiliation{Department of Chemistry, North Carolina State
University, Raleigh, North Carolina 27695-8204, U.S.A.}

\author{C. Lee}
\affiliation{Department of Chemistry, North Carolina State
University, Raleigh, North Carolina 27695-8204, U.S.A.}

\author{M.-H. Whangbo}
\affiliation{Department of Chemistry, North Carolina State
University, Raleigh, North Carolina 27695-8204, U.S.A.}

\author{R. K. Kremer}
\affiliation{Max Planck Institut f\"ur Festk\"orperforschung,
Heisenbergstr. 1, D-70569 Stuttgart, Germany}

\date{\today}

\begin{abstract}
We investigated the magnetic and structural properties of the quasi-one dimensional 3$d^1$-quantum chain system TiPO$_4$ ($J \sim$ 965 K) by magnetic susceptibility, heat capacity, ESR, x-ray diffraction, NMR measurements, and by density functional calculations. TiPO$_4$ undergoes two magnetostructural phase transitions, one at 111 K and the other at 74 K. Below 74 K, NMR detects two different $^{31}$P signals and the magnetic susceptibility vanishes, while DFT calculations evidence a bond alternation of the Ti\ldots Ti distances within each chain.  Thus, the 74~K phase transition is a spin-Peierls transition which evolves from an incommensurate phase existing
between 111~K and 74~K.
\end{abstract}

\maketitle

The discovery of high-$T_c$ superconductivity in two-dimensional oxocuprates has stimulated broad interest in the properties of low-dimensional quantum $S$=1/2 antiferromagnets.
The complex interplay between spin, charge, orbital and lattice degrees of freedom in low-dimensional systems with pronounced quantum fluctuation renders a plethora of complex and unusual ground states.\cite{Dagotto2005,Lee2008}

Most of the prominent examples of low-dimensional quantum antiferromagnets with exotic ground states contain Cu$^{2+}$ (3$d^9$, $S$ = 1/2) ions with one hole present in the $e_g$ orbitals.\cite{Enderle2005,Banks09,Law10} Compounds of early transition-metal elements with one electron in the $d$ shell are less frequently investigated.
With a 3$d^1$ electron in a high-symmetry or slightly distorted octahedral environment, the orbital degeneracy of the $t_{2g}$ states opens new degrees of freedom, with the possibility of low-energy orbital excitations and the interesting scenario of destabilization of coherent spin/orbital ordering by quantum fluctuations. 3$d^1$ systems can be easily realized in compounds containing, e.g., Ti$^{3+}$ or V$^{4+}$ cations. A paramount example is the vanadium ladder compound $\alpha$'-NaV$_2$O$_5$.\cite{Isobe1996}  At high temperatures $\alpha$'-NaV$_2$O$_5$ contains
mixed-valent vanadium cations with one electron occupying an orbital confined to the rungs of the ladder, hence constituting a quarter-filled ladder system.\cite{Smolinski1998}
Below $\sim$~34~K $\alpha$'-NaV$_2$O$_5$ undergoes charge ordering\cite{Ludecke1999} leading to a spin-gap of $\sim$~100~K indicated by a rapid drop of the magnetic susceptibility. That  was initially ascribed to a spin-Peierls transition.\cite{Isobe1996,Weiden1997}
Other prominent low-dimensional 3$d^1$ systems that have recently attracted much attention are the Mott insulators TiOX (X=Cl, Br).\cite{Beynon1993,Seidel2003,Kataev2003,Ruckamp2005,Krimmel2006} These compounds crystallizing in the FeOCl-type structure, consisting of Ti - O - X layers made up of TiO$_4$Cl$_2$ octahedra. These layers are stacked with van der Waals interactions between them.\cite{Schafer1958,Schnering1972} The magnetic susceptibility of TiOCl
reveals several unusual features, which led  to the proposal that
TiOCl may be a manifestation of a resonating valence-bond solid.\cite{Beynon1993} Subsequently, Seidel \textit{et al.} demonstrated that the high temperature susceptibility fits very well to a
$S$=1/2 Heisenberg chain model with  nearest-neighbor (nn) antiferromagnetic (afm) spin-exchange (SE) interaction of $\sim$660~K. In view of these findings and their LDA + $U$ electronic structure calculations,  Seidel \textit{et al.} concluded that TiOCl is an example of a Heisenberg chain that undergoes a spin-Peierls transition at 67~K.\cite{Seidel2003} Subsequent low temperature x-ray structure determination showed a slight dimerization of the Ti\ldots Ti distances along the $b$ direction.\cite{Shaz2005} A further
anomaly was detected at $T_{c2}$ = 95~K, which signals a  transition into
an incommensurate phase with a slight monoclinic distortion of the lattice. It was initially believed and it appears to be now generally accepted to be of continuous or higher order, however, there was
a later claim that it was first order. \cite{Krimmel2006,Hemberger2005,Schonleber2008,Zhang2008,Clancy2010}

\begin{figure}[h!]
  \includegraphics[width=9 cm]{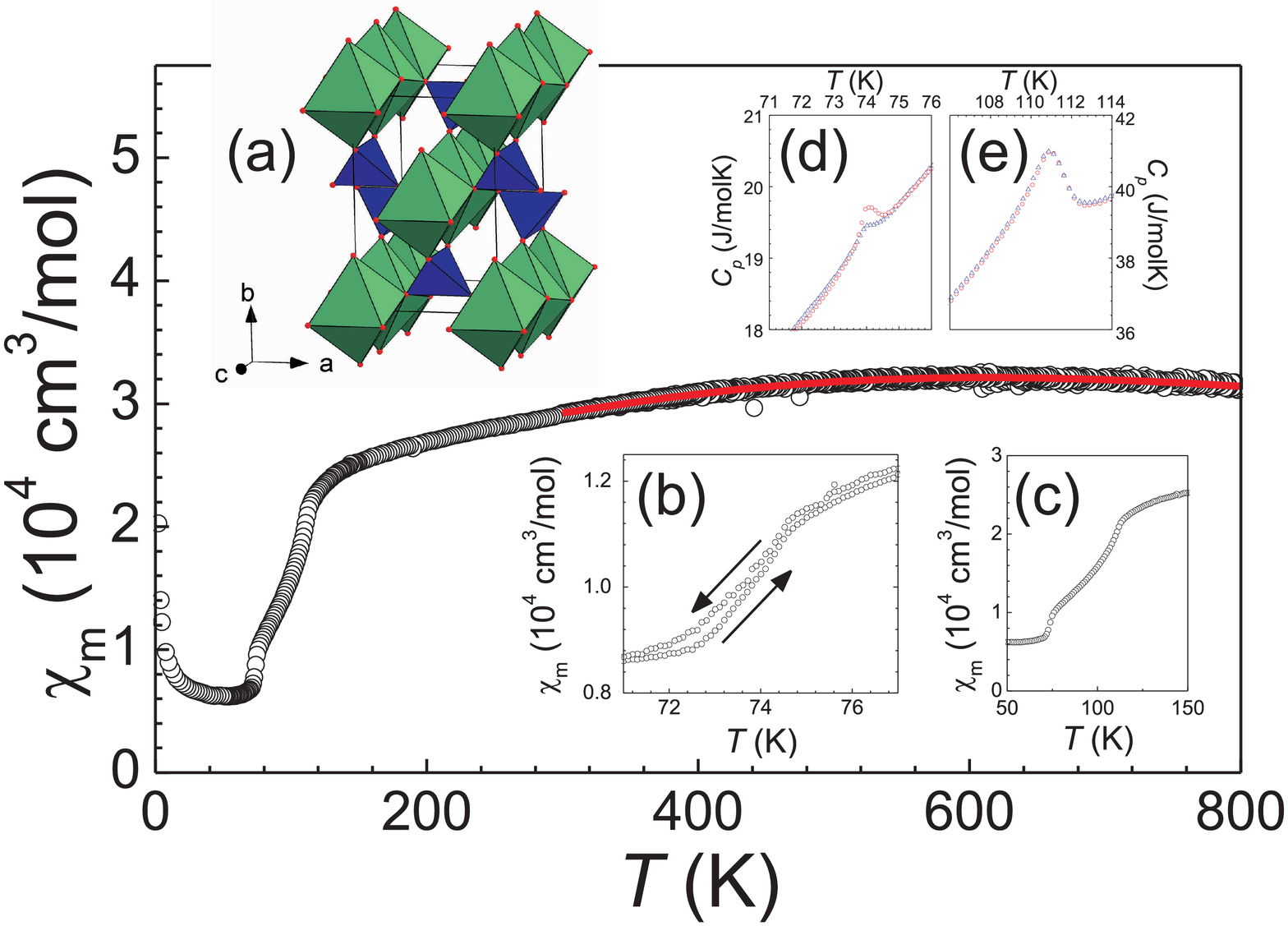}\\
  \caption{(Color online) (o) Molar magnetic susceptibility, $\chi_{\rm m}$, of TiPO$_4$ measured in a field of  1T. The (red) line is a fit to a Heisenberg chain with uniform nn afm SE interaction, see text. Insets: (a) Crystal structure of TiPO$_4$, where green and blue polyhedral represent the TiO$_6$ octahedra and PO$_4$ tetrahedra, respectively. (b, c) $\chi_{\rm m}$ in the region of the phase transitions. (d, e) Heat capacity, $C_p$, in the region of the anomalies, where the red circles and blue triangles refer to the heating and cooling data, respectively.}\label{Fig1}
\end{figure}

Here, we report the magnetic and structural properties of TiPO$_4$, which contains Ti$^{3+}$ cations and displays two magneto-structural phase transitions reminiscent of those in TiOX.
In contrast to TiOCl, however, TiPO$_4$ is a structurally one-dimensional compound
crystallizing in the CrVO$_4$ structure-type (SG: $Cmcm$) (see inset in Fig. \ref{Fig1}).\cite{Kinomura1982} The Ti$^{+3}$ ions, carrying $S$=1/2 entities, are subject to axially compressed TiO$_6$ octahedra. These share their edges to form corrugated TiO$_4$ ribbon chains along the $c$-axis with a buckling angle of 156.927(4)$^\circ$ in the \emph{a}-\emph{c} plane. The Ti$^{3+}$\ldots Ti$^{3+}$ distance at room temperature (RT) amounts to 3.1745(10)${\rm {\AA}}$ with a Ti$^{3+}$-O$^{2-}$-Ti$^{3+}$ $\angle$ of 95.484(5)$^\circ$.\cite{Glaum1992} The TiO$_4$ ribbon chains are interconnected by sharing corners with distorted PO$_4$ tetrahedra.

\begin{figure}[h!]
\includegraphics[width=8.5 cm]{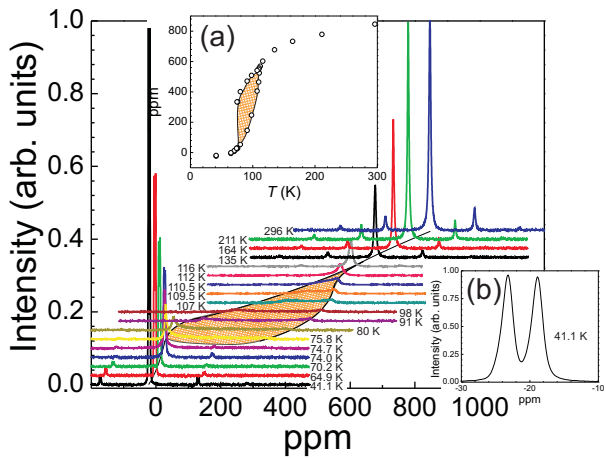}\\
\caption{(Color online) $^{31}$P MAS NMR spectra for TiPO$_4$ (temperatures  indicated). The (orange)  hashed area highlights the incommensurate  continuum. Insets: (a) Peak positions and/or boundary edges versus temperature.  (b) The 41.1~K spectrum.}\label{nmr}
\end{figure}

At high temperature the magnetic susceptibility of a polycrystalline sample (Fig. \ref{Fig1}, main panel) is characterized by a broad maximum centered at $\sim$~625~K, indicating short range afm correlations. After correction for a temperature independent offset to the susceptibility arising from diamagnetic contributions of the closed shells and van Vleck terms, the high temperature magnetic susceptibility can be described very well by  a \emph{S}=1/2 Heisenberg chain with a uniform nn afm SE interaction\cite{Johnston2000} of 965(10)~K  and a \emph{g}-factor of 1.94(3).

Below $\sim$~120~K the susceptibility reveals two subsequent magnetic phase transitions, indicated by two rapid drops of the susceptibility  at 111(1)~K and 74(0.5)~K. Finally, at lowest temperatures the susceptibility levels off to a value of 75(10)$\times$10$^{-6}$ cm$^3$/mol. At very low temperature a slight increase is seen, which we ascribe to a Curie tail due to $\sim$70 ppm of a free \emph{S}=1/2 spin entities. The anomaly at 74~K shows a thermal hysteresis with a temperature difference of $\sim$~50~mK between the heating and cooling traces while heating/cooling cycles gave identical susceptibilities for the 111~K anomaly (see Fig. \ref{Fig1} inset (b, c)).

Heat capacities collected on crystals exhibit two $\lambda$-type anomalies at 110.9(0.6)~K and 74.1(0.3)~K, with the lower temperature anomaly also showing  a thermal hysteresis, while again no hysteresis is seen for the higher temperature anomaly (see Fig. \ref{Fig1} inset (d, e)).
Angular and temperature dependent Electron Spin Resonance measurements (ESR) on single crystals revealed a single Lorentzian resonance line ($g$-factor 1.93 - 1.95) and a   linewidth decreasing linearly with temperature (50 Oe $\leq$ FWHM $\leq$ 300 Oe)   consistent with earlier findings.\cite{Kinomura1982,Glaum1996}
The integrated intensity of the ESR line mimics the temperature dependence of the magnetic susceptibility and drops to zero below 74~K.

The crystal structure of a very high quality single crystal of TiPO$_4$ was determined by x-ray single crystal diffraction measurements  at various temperatures between 293~K and 90~K.   Down to 90~K the structure was found to be identical to that reported by Glaum $\textit{et al.}$, except for small changes of the lattice parameters and the general atomic positions.\cite{Glaum1992}
As the temperature is lowered from RT to $\sim$120~K, the lattice parameters \textit{a} and \textit{b} increase
slightly but the \textit{c} parameter decreases, such that the cell volume remains almost constant.   The residual electron density, i.e. the measured electron density minus the calculated electron density (from the superposition of spherical atom densities) shows considerable residuals located within the \textit{a} - \textit{c} plane next to the Ti atoms, at an interstitial position in the ribbon chain, and at a position bisecting the \mbox{O - Ti - O} angle perpendicular to the ribbon chains.
Upon cooling there is a gradual migration of the density away from the bisecting position into the interstitial position within the ribbon chains. This change is complemented by a reduction of the distance in the \mbox{Ti - O - Ti}, nn super-exchange pathway, and  an increase of the intrachain \mbox{O - Ti - O} angle. Evidence for a structural change was not found in this temperature range, possibly due to the dynamic character of the intermediate phase as observed by NMR (see below).

Magic angle spinning (MAS)  $^{31}$P nuclear magnetic resonance (NMR) spectra (center field $\sim$~8.5 T) operating at a spinning frequency of $\sim$~25~kHz, were collected on a polycrystalline sample between $\sim$~35~K and RT. The spectra are displayed versus temperature in the main panel of Fig. \ref{nmr}. Above $\sim$~140~K we observe a single $^{31}$P symmetric NMR line accompanied by two sets of very weak symmetrically placed spinning sidebands.  Near 116~K the line becomes asymmetric and below 111~K it broadens into an asymmetric continuum limited by two boundary peaks. With decreasing temperature the continuum expands and its intensity decreases. Towards $\sim$~76~K the continuum finally washes out, whereupon its lower boundary grows into two symmetric lines
indicating the occurrence of two different P atom environments (see inset Fig. \ref{nmr} (b)). The peak positions and/or  boundary edges are shown versus temperature in Fig. \ref{nmr} (a). There are similarities between NMR
measurements of TiPO$_4$, reported here, and TiOX, reported by Saha \textit{et al.}.\cite{Saha2007}

\begin{figure}[h!]
\includegraphics[width=7 cm]{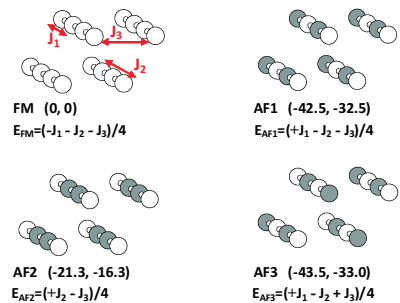}\\
\caption{(Color online) The four ordered spin configurations, FM, AF1, AF2 and AF3, used to extract the values of $J_1$, $J_2$ and $J_3$, where only the Ti$^{3+}$ ions are shown for simplicity. The up- and down-spin Ti$^{3+}$ sites are indicated by  different colors. The numbers in parenthesis (from left to right) represent the relative energies in meV per four FUs obtained from GGA+$U$ calculations with $U_{\rm eff}$ = 2 and 3~eV, respectively. The expression of the total SE energy per four FUs is also given.
}\label{exchange}
\end{figure}

We now probe the SE interactions of TiPO$_4$ by performing mapping analysis based on density functional calculations.\cite{Whangbo2003}  We consider the nn and next-nearest neighbor (nnn) intrachain SE interactions $J_1$ and $J_2$, respectively, as well as the interchain SE interaction $J_3$ (see Fig. \ref{exchange}). To evaluate $J_1$ - $J_3$, we determine the relative energies of the four ordered spin states, FM, AF1, AF2, and AF3  shown in Fig. \ref{exchange}, by density functional theory (DFT) electronic band structure calculations. Our DFT calculations employed the Vienna \textit{ab initio} simulation package (Ref. \onlinecite{Kresse1993,Kresse1996a}) with the projected augmented-wave method, the generalized gradient approximation (GGA) for the exchange and the correlation functional.\cite{Perdew1996} We used a  plane-wave cut-off energy of 400~eV, a set of 56 \textbf{k}-point irreducible Brillouin zone, and the threshold of 10$^{-5}$~eV for the self-consistent-field convergence of the total electronic energy. To account for the electron correlation associated to the Ti 3$d$ state, we performed GGA plus onsite repulsion (GGA+$U$) calculations (Ref. \onlinecite{Dudarev1998}) with an effective $U_{eff}$ = $U$ - J = 2~eV and 3~eV on Ti. The relative energies, per four formula units (FUs), of the four ordered spin states  are summarized in Fig. \ref{exchange}.

The total SE energies of the four ordered spin states  can be expressed in terms of a Heisenberg spin Hamiltonian,
$H$ = -$\sum J_{ij} \vec{{S_i}}\vec{{S_j}}$,
where $J_{\rm ij}$  is the SE interaction between the spins  $\vec{{S_i}}$ and $\vec{{S_j}}$  on the spin sites $i$ and $j$, respectively.
By applying the energy expressions obtained for spin dimers with $N$ unpaired spins per spin site (in the present case, $N$ = 1),\cite{Dai}, the total SE energies for the four configurations, per four FUs, are given in Fig. \ref{exchange}.

Thus, by mapping the relative energies of the four ordered spin configurations given in terms of the SE parameters (see Fig. \ref{exchange}) onto the corresponding relative energies obtained from the GGA+$U$ calculations, we obtained the values for the SE parameters $J_1$ - $J_3$ (see Table \ref{Table1}).\cite{Whangbo,Whangbo2003}
The results of our DFT calculations are in very good quantitative agreement with our experimental findings indicating a very large nn intrachain afm SE interaction. The nnn intrachain SE interaction is almost two orders of magnitude smaller, the interchain interaction $J_3$ amounts to 2\% of $J_1$.

\begin{table}[tbh]
\begin{ruledtabular}
\begin{tabular}{ccc}
  $J_i$ & $U$ = 2eV &  $U$ = 3eV\\
\hline
$J_1$ & -988 & -751  \\ \\
$J_2$ & -1.4 & +0.7   \\ \\
$J_3$ & -20 & -15 \\

\end{tabular}
\end{ruledtabular}
\caption[]{Values of the SE parameters $J_1$ - $J_3$ derived from the mapping analysis (in K).
\label{Table1}}
\end{table}

The low temperature MAS data of TiPO$_4$ prove a non-magnetic ground state with two distinct P atomic environments, as evidenced especially by the low temperature spectra. The chemical shifts of the $^{31}$P lines amount to $\sim$~ -20 ppm  in good agreement with what has been found for other diamagnetic orthophosphates, proving the non-magnetic character of the ground state of TiPO$_4$.\cite{Cheetham1986,Turner1986} We ascribe the 74~K phase transition in TiPO$_4$ to a spin-Peierls transition with the Ti\ldots Ti bond alternation within the Ti chains.

To probe the low temperature crystal structure of TiPO$_4$, we considered the subgroups \textit{Amm2} and \textit{Pmmn} of the RT space group $Cmcm$. By GGA calculations, we optimized the structures of TiPO$_4$ starting with the initial  settings described by \textit{Cmcm}, \textit{Amm2} and \textit{Pmmn} without symmetry constraints in order to allow the atom positions to relax freely (with a set of 28 \textbf{k}-point irreducible Brillouin zone, and the thresholds of 10$^{-5}$ eV and 0.001 eV/\AA\ for the self-consistent-field convergence of the total electronic energy and force, respectively).
The lowest-energy structure found was obtained   starting from the $Pmmn$ initial setting,  and was lower in energy by $\sim$~48~meV per formula unit than was the \textit{Cmcm} structure, and by $\sim$~32~meV per formula unit than the \textit{Amm2} initial structure. The structure relaxed from the \textit{Pmmn} initial setting  shows a dimerization in the Ti chains with alternating  Ti\ldots Ti distances of $\sim$2.9 and $\sim$3.5 \AA, which is comparable in magnitude to that observed in TiOCl.\cite{Shaz2005}  This structure also has two different  environments for the P atoms within the PO$_4$ units, which is consistent with our MAS NMR spectra.
From model calculations using the relaxed crystal structure data, we expect very weak superstructure reflections, which were not resolved  in an early neutron powder diffraction experiment.\cite{Glaum3}

The incommensurate phase seen between $\sim$~111~K and $\sim$~74~K is similar to that found for TiOX, where it has been attributed to a frustration between the spin-Peierls pairing and an elastic interchain coupling.\cite{Schonleber2008} Analogous arguments may apply for TiPO$_4$ as well. In view of the small interchain coupling the incommensurate phase between 111~K and 74~K could also be ascribed to a dynamic equilibrium between short-range-ordered dimerized segments.

In conclusion, our magnetic susceptibility, heat capacity, ESR and $^{31}$P MAS NMR measurements supported by our density functional calculations show that TiPO$_4$ undergoes a spin-Peierls transition at 74.1(0.3)~K, which is preceded by an incommensurate phase extending up to $\sim$~111~K. The thermal hysteretic behavior of these transitions is consistent with the pattern of discontinuous and continuous transitions seen for TiOCl and TiOBr.
At high temperatures the magnetic susceptibility of TiPO$_4$  is described by a $S$=1/2 Heisenberg antiferromagnetic chain with an unprecedented nn SE of $\sim$~1000~K.

 \begin{acknowledgments}
Work at NCSU by the Office of Basic Energy Sciences, Division of Materials Sciences, U. S. Department of Energy, under Grant DE-FG02-86ER45259, and also by the computing resources of the NERSC center and the HPC center of NCSU.
\end{acknowledgments}

\bibliographystyle{apsrev}

\end{document}